\documentstyle[12pt]{article}
\def\PRL #1 #2 #3{{\sl Phys. Rev. Lett.} {\bf#1} (#2) #3}
\def\NPB #1 #2 #3{{\sl Nucl. Phys.} {\bf B#1} (#2) #3}
\def\NPBFS #1 #2 #3 #4{{\sl Nucl. Phys.} {\bf B#2} [FS#1] (#3) #4}
\def\CMP #1 #2 #3{{\sl Commun. Math. Phys.} {\bf #1} (#2) #3}
\def\PRD #1 #2 #3{{\sl Phys. Rev.} {\bf D#1} (#2) #3}
\def\PLA #1 #2 #3{{\sl Phys. Lett.} {\bf #1A} (#2) #3}
\def\PLB #1 #2 #3{{\sl Phys. Lett.} {\bf #1B} (#2) #3}
\def\JMP #1 #2 #3{{\sl J. Math. Phys.} {\bf #1} (#2) #3}
\def\PTP #1 #2 #3{{\sl Prog. Theor. Phys.} {\bf #1} (#2) #3}
\def\SPTP #1 #2 #3{{\sl Suppl. Prog. Theor. Phys.} {\bf #1} (#2) #3}
\def\AoP #1 #2 #3{{\sl Ann. of Phys.} {\bf #1} (#2) #3}
\def\PNAS #1 #2 #3{{\sl Proc. Natl. Acad. Sci. USA} {\bf #1} (#2) #3}
\def\RMP #1 #2 #3{{\sl Rev. Mod. Phys.} {\bf #1} (#2) #3}
\def\PR #1 #2 #3{{\sl Phys. Reports} {\bf #1} (#2) #3}
\def\AoM #1 #2 #3{{\sl Ann. of Math.} {\bf #1} (#2) #3}
\def\UMN #1 #2 #3{{\sl Usp. Mat. Nauk} {\bf #1} (#2) #3}
\def\FAP #1 #2 #3{{\sl Funkt. Anal. Prilozheniya} {\bf #1} (#2) #3}
\def\FAaIA #1 #2 #3{{\sl Functional Analysis and Its Application} {\bf
#1} (#2) #3}
\def\BAMS #1 #2 #3{{\sl Bull. Am. Math. Soc.} {\bf #1} (#2)
#3} \def\TAMS #1 #2 #3{{\sl Trans. Am. Math. Soc.} {\bf #1} (#2) #3}
\def\InvM #1 #2 #3{{\sl Invent. Math.} {\bf #1} (#2) #3}
\def\LMP #1 #2 #3{{\sl Letters in Math. Phys.} {\bf #1} (#2) #3}
\def\IJMPA #1 #2 #3{{\sl Int. J. Mod. Phys.} {\bf A#1} (#2) #3}
\def\AdM #1 #2 #3{{\sl Advances in Math.} {\bf #1} (#2) #3}
\def\RMaP #1 #2 #3{{\sl Reports on Math. Phys.} {\bf #1} (#2) #3}
\def\IJM #1 #2 #3{{\sl Ill. J. Math.} {\bf #1} (#2) #3}
\def\APP #1 #2 #3{{\sl Acta Phys. Polon.} {\bf #1} (#2) #3}
\def\TMP #1 #2 #3{{\sl Theor. Mat. Phys.} {\bf #1} (#2) #3}
\def\JPA #1 #2 #3{{\sl J. Physics} {\bf A#1} (#2) #3}
\def\JSM #1 #2 #3{{\sl J. Soviet Math.} {\bf #1} (#2) #3}
\def\MPLA #1 #2 #3{{\sl Mod. Phys. Lett.} {\bf A#1} (#2) #3}
\def\JETP #1 #2 #3{{\sl Sov. Phys. JETP} {\bf #1} (#2) #3}
\def\JETPL #1 #2 #3{{\sl  Sov. Phys. JETP Lett.} {\bf #1} (#2) #3}
\def\PHSA #1 #2 #3{{\sl Physica} {\bf A#1} (#2) #3}
\def\CQG #1 #2 #3{{\sl Class. Quantum Grav.} {\bf #1} (#2) #3}
\def\SJNP #1 #2 #3{{\sl Sov. J. Nucl. Phys. (Yadern.Fiz.)} {\bf #1} (#2) #3}
\def\a{\alpha}

\def\be{\begin{equation}}\def\ee{\end{equation}}

\newcommand{\p}[1]{(\ref{#1})}
\begin{document}
\renewcommand{\thefootnote}{\arabic{footnote}}
\begin{flushright}
HUB-EP-98/36 \\
DFPB 98/TH/31\\
hep--th/9806175
\end{flushright}

\vspace{1truecm}
\begin{center}
{\large\bf 
Coupling of M--branes in M--theory}\footnote{Talk given at the Sixth 
International Symposium on Particles, Strings and Cosmology (PASCOS 98), 
Boston, March 22--29, 1998}

\vspace{1cm}
{Dmitri Sorokin}\footnote{
Alexander von Humboldt Fellow \\
{}~~~On leave from Kharkov Institute of Physics and
Technology, Kharkov, 310108, Ukraine} 

\vspace{1cm}
Humboldt-Universit\"at zu Berlin\\
Mathematisch-Naturwissenshaftliche Fakultat\\
Institut f\"ur Physik\\
Invalidenstrasse 110, D-10115 Berlin

\bigskip
and

\bigskip
Universit\`a Degli Studi Di Padova\\
Dipartimento Di Fisica ``Galileo Galilei''\\
ed INFN, Sezione Di Padova\\
Via F. Marzolo, 8, 35131 Padova, Italia

\vskip .2in
{\bf Abstract}
\end{center}
Coupling of a membrane and a five-brane to the bosonic sector of $D=11$
supergravity is considered. The five--brane is a dyonic object which
carries both an electric and a magnetic charge of the $D=11$ three-form
gauge field $A^3$, and it couples not only to $A^3$ but also 
minimally couples to a six-form field $A^6$ dual to $A^3$. 
This implies that the 5-brane should
more naturally couple to a version of $D=11$ supergravity where
both gauge fields are present in a duality-symmetric fashion.
We demonstrate how an action of duality-symmetric $D=11$ 
supergravity looks like,
couple it to the five-brane and then reduce the resulting system 
to an action, which describes an interaction of the 5-brane with 
the standard $D=11$ supergavity. 

\newpage
In this talk I would like to present results 
obtained in collaboration with Igor Bandos and Nathan Berkovits
\cite{bbs} on studying the coupling of D=11 supergravity 
to a membrane and a 5--brane, the extended objects which constitute 
a part of what is understood under the name of M--theory. 
These results generalize and complete earlier work on this subject 
done by de Alwis \cite{alwis}, and, in fact, are an extension to 
M--theory of a classical work by Dirac \cite{dirac} on coupling 
monopoles to electromagnetic fields.

The problem is to find an action which would describe $D=11$ supergravity and
M--branes as a closed dynamical system. The membrane and the 5--brane are 
charged objects with respect to a three--form gauge field $A^{(3)}_{MNP}(x)$
$(M,N,P=0,1,...,10)$ which, together with a graviton field $g_{MN}(x)$
and a gravitino field $\Psi^\alpha_M(x)$ ($\alpha=1,...,32)$ form 
the supermultiplet of $D=11$ supergravity. The action of this theory 
was constructed twenty years ago by Cremmer, Julia and Scherk \cite{cjs} 
and has the following form
\begin{equation}\label{1}
{\cal S}_{CJS}=-\int d^{11}x 
{{\sqrt{-g_{11}}}\over{2^. 4!}}F^{(4)}_{M_1...M_4}F^{(4)M_1...M_4}
+\int {1\over 3}A^{(3)}\wedge F^{(4)}\wedge F^{(4)}
\end{equation}
$$
+\int d^{11}x{\sqrt{-g_{11}}}\big ({1\over 4}R -{{i}\over
2}\bar\Psi_{M}\Gamma^{MNP}D_{N}\Psi_{P}\big )~+~S_{int}.
$$
It contains the kinetic term of the $A^{(3)}$--field constructed from its
four--form field strength $F^{(4)}=dA^{(3)}$, and the Chern--Simons term,
the Einstein action for the graviton $g_{MN}(x)$ ($g_{11}=\det g_{MN})$, 
the kinetic term for the 
gravitino $\Psi^\a_M$ and terms $S_{int}$ describing self--interaction of the
$D=11$ supermultiplet. Since below, we will deal with the bosonic part, 
and, in particular, with the field $A^{(3)}$ part of the action, $S_{int}$ 
has not been written in an explicit form. I.e. we shall put 
$\Psi^\alpha_M$ to zero everywhere. 
A reason for this is that a consistent supersymmetric description 
of $D=11$ supergravity and super--M--branes as a closed interacting 
system is still lacking. We shall return to the discussion of this 
problem in Conclusion. As to the bosonic sector, the problem of 
coupling has been solved.

$D=11$ supergravity described by the action \p{1} can be regarded as a 
``free''
supergravity theory, since all of its fields are of a supergauge or 
geometrical nature, and the right hand side of its field equations 
does not contain any matter sources. 
And for a long period of time it was believed that no matter exists in 
eleven dimensions except for supergravity itself.
Now we know that $D=11$ matter does exist in a form of membranes, 
5--branes and waves ($D=11$ superparticles) (see, for example, 
\cite{pkt} for a recent review). 
The latter we shall put aside since M--waves do not carry three--form 
field charges being the main concern of the present discussion.

A membrane can be easily coupled to the $A^{(3)}$--field  \cite{bst}
 by adding 
to its worldvolume action a corresponding minimal--coupling term:
\begin{equation}\label{2}
S=-T_2\int_{{\cal M}_3} d^3 z \sqrt{-det g_{mn}}-
T_2 \int_{{\cal M}_3} dz^{m_1}\wedge dz^{m_2}
\wedge dz^{m_3} A_{m_1 m_2
m_3}^{(3)} (x(z)) \quad (m=0,1,2),
\end{equation}
where the first term in the membrane action is an integral over worldvolume
 ${\cal M}_3$ of the membrane, whose metric $g_{mn}\equiv
{{\partial x^{P}}\over{\partial
z_m}}g_{{}_{PQ}}(x(z))
{{\partial x^{Q}}\over{\partial z_n}}
$ is induced by embedding into $D=11$ curved space--time, 
the $D=11$ three--form $A^{(3)}$
is pulled back into ${\cal M}_3$, and $T_2$ is a membrane tension 
associated with a membrane's ``electric'' charge 
which can be regarded as a source for the $A^{(3)}$--field. 
Indeed, if we consider the sum of the $D=11$ supergravity action \p{1} 
(with $\Psi^\alpha_m=0$) and the membrane action \p{2}, 
and vary the full action with respect to $A^{(3)}$, 
we get the gauge--field equation of motion which contains the membrane 
current on its right--hand side:
\begin{equation}\label{3}
d{}^* F^{(4)}+F^{(4)}\wedge F^{(4)}={~}^*J^{(3)},
\qquad  d{}F^{(4)}=0.
\end{equation}
where the $A^{(3)}$--field equations are written in terms of differential 
forms, ${}^*$ denotes the $D=11$ Hodge--duality operation, and
\begin{equation}\label{41}
J^{(3)MNP}(x)={T_2\over{\sqrt{-g_{11}}}}\int_{{\cal M}_3}
d\hat x^{M}\wedge d\hat x^{N} \wedge d\hat x^{P} \delta(x-{\hat
x}(z))
\end{equation}
is the membrane current. 
Together with the equation of motion eq. \p{3} also contains 
the Bianchi identity for $F^{(4)}$ which implies that the latter 
is an external derivative of $A^{(3)}$. 

Thus, we see that the coupling of a membrane to the bosonic sector of $D=11$
supergravity causes no problem. 
The situation with 5--brane coupling is much more complicated.
The 5--brane is a dyonic object, since it couples to both the $A^{(3)}$
field and its dual six--form field $A^{(6)}$.

On--shell duality between $A^{(3)}$ and $A^{(6)}$ is established by 
relating their field strengths in such a way that for the field $A^{(6)}$ 
the equation of motion and the Bianchi identity in \p{3} 
exchange their roles. 
In the absence of membrane sources (i.e. when $J^{(3)}=0$) 
the duality relation is
\begin{equation}\label{5}
F^{(7)}\equiv dA^{(6)}-A^{(3)}\wedge F^{(4)}= {}^* F^{(4)},
\qquad F^{(4)}= -{}^* F^{(7)},
\end{equation}
where $F^{(7)}$ is the  generalized seven--form field strength of $A^{(6)}$.
 From this duality relation it follows that the 
 Bianchi identity and the equation of motion of the field $F^{(7)}$ 
have, respectively, the 
following form:
\begin{equation}\label{6} 
d{}F^{(7)}+F^{(4)}\wedge F^{(4)}=0, \qquad
 d{}^* F^{(7)}=0.
\end{equation}
These are nothing but the original equations \p{3} (with $J^{(3)}=0$) 
rewritten
in a form which explicitly includes the dual field $A^{(6)}$.

A Wess--Zumino part of the 5--brane worldvolume action, describing
minimal coupling of the 5--brane to the $A^{(3)}$ and $A^{(6)}$ field, 
was constructed by Aharony \cite{aha} using a requirement that the 
Wess--Zumino term must respect local gauge symmetries 
\begin{equation}\label{g}
\delta A^{(3)}=d\phi^{(2)}, \qquad \delta A^{(6)}=d\phi^{(5)}+\phi^{(2)}\wedge 
F^{(4)}
\end{equation}
of the duality 
condition \p{5}. The Wess--Zumino action looks as follows:
\begin{equation}\label{7}
S_{WZ}=-{T_5\over 2}\int_{{\cal M}_6}\left[A^{(6)}+ dB^{(2)}(y) \wedge
A^{(3)}\right],
\end{equation}
where $A^{(6)}(x(y))$ and $A^{(3)}(x(y))$ are the pullbacks 
into the 5-brane worldvolume ${\cal M}_6$ of the 6--form 
and the 3--form D=11 gauge field, $T_5$ is a 5-brane coupling constant 
with respect to these fields, and
$B^{(2)}_{mn}(y)$ $(m,n=0,1...,5)$ is a two--rank gauge field 
living in the worldvolume of the 5--brane.

Supersymmetry in the 5--brane worldvolume induced by its embedding 
into $D=11$ superspace requires that on the mass shell the field strength 
of $B^{(2)}$ must satisfy a self--duality condition. 
At a linearized level this condition is the standard one:
$$
H^{(3)}_{m_1m_2m_3}={}^*H^{(3)}_{m_1m_2m_3}={{\sqrt{-g_{6}}}\over 6}
\epsilon_{m_1m_2m_3n_1n_2n_3}H^{(3)n_1n_2n_3},
$$
where 
\begin{equation}\label{71}
H^{(3)}=dB^{(2)}-A^{(3)}
\end{equation}
is an extended field--strength of 
$B^{(2)}$ containing the pullback of the $D=11$ 3--form field, and
$g_{6}(x(y))$ is the determinant of an ${\cal M}_6$--metric 
$g_{mn}$ induced by embedding into $D=11$.

Due to the presence of this self--dual field $B^{(2)}$ in the 
five--brane worldvolume the complete action for the five--brane remained 
an open problem until an essential progress has been made by
 Perry and Schwarz \cite{ps} and in \cite{pst} in understanding how 
the two--form field is incorporated into the action, and the complete
$\kappa$--symmetric action for the 5--brane propagating in D=11
superspace was constructed independently in \cite{m51} and \cite{m52}.

In addition to the Wess--Zumino term \p{7} this action has another
 two terms which, being restricted to the bosonic  sector, 
have the following Born--Infeld--like form:
\begin{equation}\label{8}
S_{{\cal M}_6}=
T_5\int_{{\cal M}_6} d^6 y \big[-\sqrt{-det(g_{mn} +{}^*{H}_{mnp}v^p)}
+\sqrt{-g_{6}}{1\over {4}}
v_l{~}^*H^{lmn}H_{mnp}v^p\big],
\end{equation}
where the vector 
$v_p={{\partial_p a(y)}\over {\sqrt{-(\partial a)^2}}}$ is a
normalized derivative of a scalar field $a(y)$. This scalar field 
ensures worldvolume covariance of the construction. It is completely 
auxiliary and can be gauge fixed by use of an available local symmetry  
in such a way that $\partial_p a(y)$ becomes a constat vector. 
Then manifest d=6 invariance of the self--dual action is lost
\footnote{In parallel to the action formulation a superembedding approach 
\cite{bpstv}
to derive 5--brane equations of motion has been developed in \cite{hsw}.
The 5--brane equations obtained with this method are equivalent \cite{eq}
to the equations which follow from the action \p{8}+\p{7}.}.

As a main consequence, the action
\p{7}+\p{8} yields the 
equation of motion of the field $B^{(2)}$ which at a linearized level 
reduces to the self--duality condition $H^{(3)}={}^*H^{(3)}$ 
(see Refs. \cite{pst,m51,m52} for details).

Because of the direct coupling of the 5--brane to the 6--form dual field
$A^{(6)}$ in the Wess--Zumino term \p{7}, we cannot just take a sum of the 
5--brane action \p{7} + \p{8} and the standard supergravity action \p{1} for a
consistent description of a closed supergravity -- 5--brane system, since 
the supergravity action does not contain $A^{(6)}$. 

To couple the standard formulation of $D=11$ supergravity to a 5--brane,
one should replace the $A^{(6)}$ field in the Wess--Zumino term with 
another (nonminimal) term containing the $A^{(3)}$ field, but, 
{\it a priori}, it is difficult to guess the form of such a term. 
So to solve the problem we choose a way around. 
The strategy is the following \cite{alwis,bbs}:
i) to construct a new formulation of $D=11$ supergravity, 
which contains both the $A^{(3)}$ and $A^{(6)}$--field in a symmetric way, 
ii) to couple this duality--symmetric action to the 5--brane action, 
iii) and finally to eliminate the field $A^{(6)}$ from the system by 
using an algebraic part of its duality relation \p{5} with $A^{(3)}$. 
At the end one should get a new form of the 
5-brane Wess--Zumino term which does not contain $A^{(6)}$.

This programme has been fulfilled in \cite{bbs}, where a complete, 
locally supersymmetric, formulation of $D=11$ supergravity with both 
gauge fields was constructed, and its bosonic sector was coupled to the 
5--brane.

A relevant gauge--field part of the duality--symmetric D=11 supergravity 
action looks as follows:
\begin{equation}\label{9}
S=-\int d^{11}x~\sqrt{-g_{11}}[{1\over{4^.
4!}}F^{(4)}_{M_1...M_4}F^{(4)M_1...M_4} +{1\over{4^.
7!}}F^{(7)}_{M_1...M_7}F^{(7)M_1...M_7}\big ] 
+\int {1\over 6}F^{(7)}\wedge F^{(4)}
\end{equation}
$$ -\int d^{11}x~\sqrt{-g_{11}}\big[{1\over{4^.  3!}} v^{P}
{\cal F}^{(4)}_{PM_1M_2M_3} {\cal
F}^{(4)QM_1M_2M_3}v_{Q}
+{1\over{4^. 6!}}v^{P}{}^*{\cal F}^{(4)}_{PM_1...M_6}{}^*{\cal
F}^{(4)QM_1...M_6}v_{Q}].$$
It contains the Maxwell terms for both the $A^{(3)}$ and the $A^{(6)}$ 
field, the
gauge-invariant term $F^{(7)}\wedge F^{(4)}$, which coincides with the 
Chern--Simons term
$-A^{(3)}\wedge F^{(4)}\wedge F^{(4)}$ up to a total derivative, and
two additional terms, which contain the 4-form tensor
$$
{\cal F}^{(4)}=F^{(4)}+{}^*F^{(7)}
$$
and its 7--form dual ${}^*{\cal F}^{(4)}$. These last two terms contain
also a vector $v_M(x)={{\partial_Ma(x)}\over{\sqrt{-(\partial a)^2}}}$ 
being a normalized derivative of an auxiliary scalar field $a(x)$. 
It turns out that for consistent coupling this duality--symmetric 
action to the 5--brane, the worldvolume auxiliary field $a(y)$ in \p{8} 
must be the pullback of the $D=11$ field $a(x)$, i.e. $a(y)=a(x(y))$.

The action \p{9} produces the duality condition \p{5} for $A^{(3)}$ and
$A^{(6)}$ as a consequence of their equations of motion. A local symmetry of
\p{9} which ensures the on-shell duality relation is given by the following 
transformations
\begin{equation}\label{10}
\delta A^{(3)}=da\wedge \varphi^{(2)}(x), \qquad
\delta A^{(6)}=da\wedge \varphi^{(5)}(x)+ \delta A^{(3)}\wedge A^{(3)},
\qquad \delta a=0,
\end{equation}
where $\varphi^{(2)}(x)$ and $\varphi^{(6)}(x)$ are, respectively a 2--form 
and a 3--form parameter.

The five--brane action \p{7} + \p{8} possesses an analogous symmetry under
\begin{equation}\label{11}
\delta B^{(2)}(y)=da(y)\wedge \varphi^{(1)}(y), \qquad \delta a=0.
\end{equation}

We must respect the symmetries \p{10} and \p{11} 
when couple the $D=11$ action and the 5--brane action together. 
And this is a reason why the worldvolume scalar field must be the 
pullback of the $D=11$ auxiliary scalar. 
But this is not the only thing to do. 
For the coupling to be consistent with the symmetries
\p{10} and \p{11} one should further modify the $D=11$ action, 
and here a classical work of Dirac \cite{dirac} on the generalization 
of the Maxwell action in the presence of monopoles provides us with a 
way of reaching the goal. 
The Dirac idea consists in the following. A gauge field associated with
magnetically charged objects, such
as four--dimensional monopoles, or the $D=11$ 5--brane, intrinsically 
possesses 
a nontrivial topological structure, which can be described by an unobserved, 
so called Dirac's, string stemmed from the point--like monopole. It appears
in the theory to solve the problem of modified Bianchi identities for the
gauge field strength. The Bianchi identities 
acquire a magnetic current on their right 
hand side. For instance, the coupling of a D=4 Maxwell field to a point--like 
dyon with an electric charge $e$ and a magnetic charge $g$ is described by 
the following Maxwell equations
\begin{equation}\label{4}
\partial_m {\hat F}^{mn}=ej^n,  \qquad 
{1\over 2}\partial_m \epsilon^{mnpq} {\hat F}_{pq}=
gj^n \qquad (m,n=0,1,2,3).
\end{equation}
where $j^n(x)=\int d{\hat x}(\tau)\delta(x-{\hat x}(\tau))$ is 
the dyon current
and $\tau$ is its proper time.
Due to the magnetic source $d\hat F\not = 0$, and hence, 
the gauge field strength
is not anymore just an external derivative of a vector potential $A_m$, but a 
tensor of the form ${\hat F}=dA-{}^*G$, where a two-rank tensor $G_{mn}(x)$ 
is related to the magnetic current through the condition $d{}^*G=-g{~}^*j$.
A solution to this equation is
\begin{equation}\label{120}
G^{mn}=-g\int_{{\cal M}_2} d\hat x^m\wedge d\hat x^n\delta(x-\hat x(z)), 
\qquad (z~\epsilon~{\cal M}_2),
\end{equation} 
where the integral is taken over a two-dimensional worldsheet ${\cal M}_2$
of the Dirac string stemmed from the dyon to infinity. The boundary of 
${\cal M}_2$ is the dyon worldline. A physical meaning of the Dirac string
is that a magnetic flux runs out of or towards
the magnetic particle to infinity along
the string.

The action which yields the Maxwell equations \p{4} and equations of motion 
of the dyonic particle is 
\begin{equation}\label{61}
S=-\int d^4 x {1\over 4}{\hat F}_{mn}{\hat F}^{mn}
-\int d\tau \big (m \sqrt{-{\dot x}^m{\dot x}_n}
+eA_m{{\partial x^m(\tau)}\over{\partial \tau}}\big).
\end{equation}
Note that the dyon minimally couples to $A_m$ 
only through its electric charge, 
since the potential dual to $A_m(x)$ has not been involved into the
construction of the action \p{61}. 
The magnetic interaction of the dyon is only through the Dirac string.
In a duality--symmetric form the Dirac action was reformulated in 
\cite{ht,mb}.

Now let us apply the same idea to the dyonic 5--brane. In this case the 
analogue of the Dirac string is a six--dimensional object (a Dirac 6--brane)
which is ended on the 5--brane. The Dirac 6-brane is described by a D=11
7--form having the following properties:
\begin{equation}\label{12}
G^{(7)M_1...M_7}(x)=
{T_5\over{\sqrt{-g_{11}}}}
\int_{{\cal M}_7}d{\hat x}^{M_1}\wedge...\wedge d{\hat x}^{M_7}
\delta(x-{\hat x}(z)), \quad  d{}^*G^{(7)}\equiv {~}^*J^{(6)},\quad
(z~\epsilon ~{\cal M}_7)
\end{equation}
$$
J^{(6)M_1...M_6}(x)=
{T_5\over\sqrt{-g_{11}}}\int_{{\cal M}_6}d{\hat x}^{M_1}\wedge ...
\wedge d{\hat x}^{M_6}
\delta(x-{\hat x}(y)), 
$$
where the integral is taken over a 7--dimensional worldvolume ${\cal M}_7$
 of the Dirac 6--brane whose boundary is the 5--brane worldvolume 
$\partial{\cal M}_7={\cal M}_6$, and $J^{(6)}(x)$ is the 5--brane current.

As we have learned from the example of the $D=4$ dyon, 
the prescription of Dirac for
coupling magnetically charged $p$--objects is to extend the
field--strengths of gauge fields with corresponding $(p+2)$--forms.
In the $D=11$ case the proper extension is:
\begin{equation}\label{13}
{\hat F}^{(4)}=dA^{(3)}-{}^*G^{(7)}, \qquad 
{\hat F}^{(7)}=dA^{(6)}-A^{(3)}\wedge dA^{(3)} -H^{(3)}\wedge{}^*G^{(7)},
\end{equation}
where the $H^{(3)}$ field strength \p{71} of the 5--brane should be 
understood 
as formally extended to a 3--form in $D=11$.
 
The action describing the coupling of the $A^{(3)}$ and $A^{(6)}$ field to 
the 5--brane has the following form:
\begin{equation}\label{14}
S=S_{{\cal M}_6}+S_{D11}({\hat F})-{1\over 6}\int A^{(3)}\wedge d A^{(3)}
\wedge d A^{(3)}-{1\over 2}\int H^{(3)}\wedge {\hat F}^{(4)}\wedge{}^*G^{(7)},
\end{equation}
where $S_{{\cal M}_6}$ is the 5--brane action \p{7} + \p{8} and 
$S_{D11}({\hat F})$ is the  part of the duality symmetric 
action \p{9} quadratic in $F$ where the field strengths are extended 
as in eq. \p{13}.
Note also the appearance of a new, the last term in the action \p{14}.

This action produces the duality relation between the extended field strengths
\p{13}:
\begin{equation}\label{15}
{\hat F}^{(7)}={}^*{\hat F}^{(4)}, \qquad {\hat F}^{(4)}=-{}^*{\hat F}^{(7)}.
\end{equation}
Taking the external derivative of these equations, and taking into account 
\p{12}, we get the equations of motion of the gauge fields $A^{(3)}$ and 
$A^{(6)}$ with the 5--brane source on their right hand side:
\begin{equation}\label{16}
d{}^*{\hat F}^{(4)}+dA^{(3)}\wedge{\hat F}^{(4)}=H^{(3)}\wedge{}^*J^{(6)},
\qquad d{}^*{\hat F}^{(7)}\equiv -d{\hat F}^{(4)}={~}^*J^{(6)}.
\end{equation}
We observe that by the use of the duality relation \p{15} we can exclude
the 6--form field from the eqs. \p{16}, which then describe the coupling of 
the 5--brane solely to the $A^{(3)}$--field.
It turns out that the 6--form field can also be eliminated directly from the 
action \p{4} in a consistent way, as shown in \cite{bbs}. As a result we get 
the coupling of the 5--brane to the bosonic sector of the standard version of
$D=11$ supergravity described by the action
\begin{equation}\label{17}
S=\int d^{11}x \sqrt{-g_{11}}[{1\over 4}R-
{1\over{2^.4!}}\hat F^{(4)}_{M_1...M_4}
\hat F^{(4)M_1...M_4}]
+\int {1\over 3}A^{(3)}\wedge dA^{(3)}\wedge dA^{(3)}
\end{equation}
$$
-T_5\int_{{\cal M}_6} d^6 y \big[\sqrt{-det(g_{mn} + {}^*{H}_{mnp}v^p)}
+{{\sqrt{-g_6}}\over {4}}
v_l{~}^*H^{lmn}H_{mnp}v^p\big]
$$
$$ -
{T_5\over 2}\int_{{\cal M}_6} dB^{(2)} \wedge
A^{(3)} + {1\over 2}\int A^{(3)}\wedge d A^{(3)}\wedge {}^*G^{(7)}.
$$
Note that the last $D=11$ term in this action replaces the minimal 
$A^{(6)}$--coupling term in the Wess--Zumino part \p{7} 
of the original worldvolume  5--brane action.
Using the definition of $G^{(7)}$ \p{12}, one can easily check that \p{17} is 
invariant under gauge transformations
$\delta A^{(3)}= d\phi^{(2)}(x)$, $\delta B^{(2)}= \phi^{(2)}(x(y))$ 
up to a total derivative.

In conclusion, we have carried out the coupling of the bosonic sector
of $D=11$ supergravity action to the membrane and the 5--brane effective 
action. This coupling also admits an extension to the case when a membrane 
is ended on a 5--brane \cite{alwis,bbs}. The membrane action
\p{2} then aquires an additional term
$$
T_2\int_{\partial{\cal M}_3}B^{(2)}(y(z))=
T_2\int_{{\cal M}_6}B^{(2)}(y)\wedge{}^*j^{(2)}(y),
$$
required by gauge symmetry conservation at the boundary 
$(\delta A^{(3)}=d\phi^{(2)},~\delta B^{(2)}=\phi^{(2)})$.
$$
j^{(2)mn}(y)={1\over{\sqrt{-g_6}}}\int_{\partial{\cal M}_3}d\hat y^m(z)
\wedge d\hat y^n(z)\delta(y-\hat y(z))\quad (m,n=0,1,...,5).
$$
At the same time, to preserve the local symmetries \p{11} of the five--brane
action the field strength $H^{(3)}$ should be extended to
$H^{(3)}-{T^2\over T^5}{}^*G^{(3)}$, where $G^{(3)}$ satisfies the equation
$d{}^*G^{(3)}\equiv j^{(2)}$ and describes a Dirac membrane inside the 
five--brane. It is tempting to identify this Dirac membrane with the
physical membrane which does not end but penetrates the five--brane.

As discussed in the literature \cite{alwis,anomaly}, the action of this 
kind can be
useful for understanding anomalies associated with the presence 
of the 5--branes in M--theory. Note that an anomaly inflow from the 
$D=11$ bulk runs to the 5--brane along a 6--brane.

As a subject of further study, one may try to analyse the possibility of 
extending these bosonic actions to locally supersymmetric actions. This 
problem is general for all super--p--branes in supergravity backgrounds and 
is connected with the following facts. Kappa--symmetric and 
target--space supersymmetric actions for the superbranes are naturally 
constructed in curved target {\it superspaces}. An important point is that 
$\kappa$--symmetry requires background supergravity superfields to
satisfy superfield constraints, which (in most of the cases) are equivalent 
to free supergravity equations without super--p--brane sources. 
Thus, to make a progress (if any) towards solving the problem of 
supersymmetric coupling one should either look for a generalization of 
superfield supergravity constraints which would include super--p--brane 
currents, or to construct worldvolume actions for the p--branes which
would couple only to physical component fields of corresponding supergravity.
In the latter case one may hope that the consistency of such actions will 
require less severe restrictions on the background supergravity fields 
than $\kappa$--symmetry of the superfield actions.

\bigskip
\bigskip
\noindent
{\bf Acknowledgements}. 
The author would like to thank Paolo Pasti and Mario Tonin for kind
hospitality at Padova University and useful discussions. This work was 
partially supported by the INTAS Grants N 93--493--ext and N96--0308.

\end{document}